\documentclass[conference]{IEEEtran}
\usepackage{subcaption} 
\IEEEoverridecommandlockouts
\usepackage{amsmath}
\usepackage{cite}
\usepackage{soul}
\usepackage{hyperref}
\sethlcolor{green}
\usepackage[table]{xcolor}
\usepackage{multirow}
\usepackage{graphicx}
\usepackage{amsmath,amssymb,amsfonts}
\usepackage{algorithmic}
\usepackage{subcaption}  

\usepackage{textcomp}
\usepackage{xcolor}
\def\BibTeX{{\rm B\kern-.05em{\sc i\kern-.025em b}\kern-.08em
    T\kern-.1667em\lower.7ex\hbox{E}\kern-.125emX}}

\begin{document}

\title{Enhanced
 Multimodal Hate Video Detection
via Channel-wise and Modality-wise Fusion\\
}

\author{%
    \parbox{\linewidth}{\centering
         Yinghui Zhang$^1$, Tailin Chen$^1$, Yuchen Zhang$^{1,2}$, Zeyu Fu$^1$ \\
        $^1$Department of Computer Science, University of Exeter, UK\\
         $^2$Institute for Analytics and Data Science, University of Essex, UK\\
        \small \{yz949,t.chen2,z.fu\}@exeter.ac.uk,
        \small yuchen.zhang@essex.ac.uk
    }
}
\maketitle

\begin{abstract}
The rapid rise of video content on platforms such as TikTok and YouTube has transformed information dissemination, but it has also facilitated the spread of harmful content, particularly hate videos. Despite significant efforts to combat hate speech, detecting these videos remains challenging due to their often implicit nature. Current detection methods primarily rely on unimodal approaches, which inadequately capture the complementary features across different modalities. While multimodal techniques offer a broader perspective, many fail to effectively integrate temporal dynamics and modality-wise interactions essential for identifying nuanced hate content.
In this paper, we present CMFusion, an enhanced multimodal hate video detection model utilizing a novel Channel-wise and Modality-wise Fusion Mechanism. CMFusion first extracts features from text, audio, and video modalities using pre-trained models and then incorporates a temporal cross-attention mechanism to capture dependencies between video and audio streams. The learned features are then processed by channel-wise and modality-wise fusion modules to obtain informative representations of videos. Our extensive experiments on a real-world dataset demonstrate that CMFusion significantly outperforms five widely used baselines in terms of accuracy, precision, recall, and F1 score. Comprehensive ablation studies and parameter analyses further validate our design choices, highlighting the model's effectiveness in detecting hate videos. The source codes will be made publicly available at  \url{https://github.com/EvelynZ10/cmfusion}.
\end{abstract}


\section{Introduction}

The proliferation of the Internet and social media has fundamentally transformed how information is disseminated, with video content emerging as a powerful and dominant medium. In 2021 alone, TikTok saw approximately 3.2 billion videos uploaded\footnote{TikTok Community Guidelines, https://www.tiktok.com/transparency/en-us/community-guidelines-enforcement-2021-4/.}, while YouTube experienced uploads at a staggering rate of 500 hours per minute by early 2022\footnote{UGC 
Statistics Facts, https://www.statista.com/topics/1716/user-generated-content/\#topicOverview.}. This massive influx of videos, viewed and shared by millions globally, has not only revolutionized communication but also facilitated the spread of harmful content, particularly hate videos\footnote{For clarity, the terms \textit{hate videos},  \textit{hate content}, and  \textit{hateful content} are used interchangeably in this paper to refer specifically to videos that contain hate speech.}. 
Hate speech, as defined by UNESCO, refers to any speech and behaviour that may incite violence, discrimination, or hostility based on race, gender, religion, or other social attributes\footnote{United Nations Educational, Scientific and Cultural Organization, https://www.unesco.org/en/countering-hate-speech/need-know.}. 
Given that hateful content infringes upon individuals' dignity and rights, while also potentially exacerbating social conflicts, both social media platforms and the academic community have made substantial efforts to combat hate videos. However, these measures have yielded limited success \cite{mathew2020hate,chandrasekharan2017you,lin2021early}.
The detection of hate videos remains particularly challenging due to their implicit and often concealed nature. Hate videos may not always be overtly aggressive but can be embedded in subtle forms such as satire, irony, or coded language. Additionally, the use of visual and auditory cues, memes, and layered messaging further complicates the identification process.
\begin{figure}[!tbp]  
    \centering
    \begin{subfigure}{0.49\linewidth}  
        \centering
        \includegraphics[width=\linewidth]{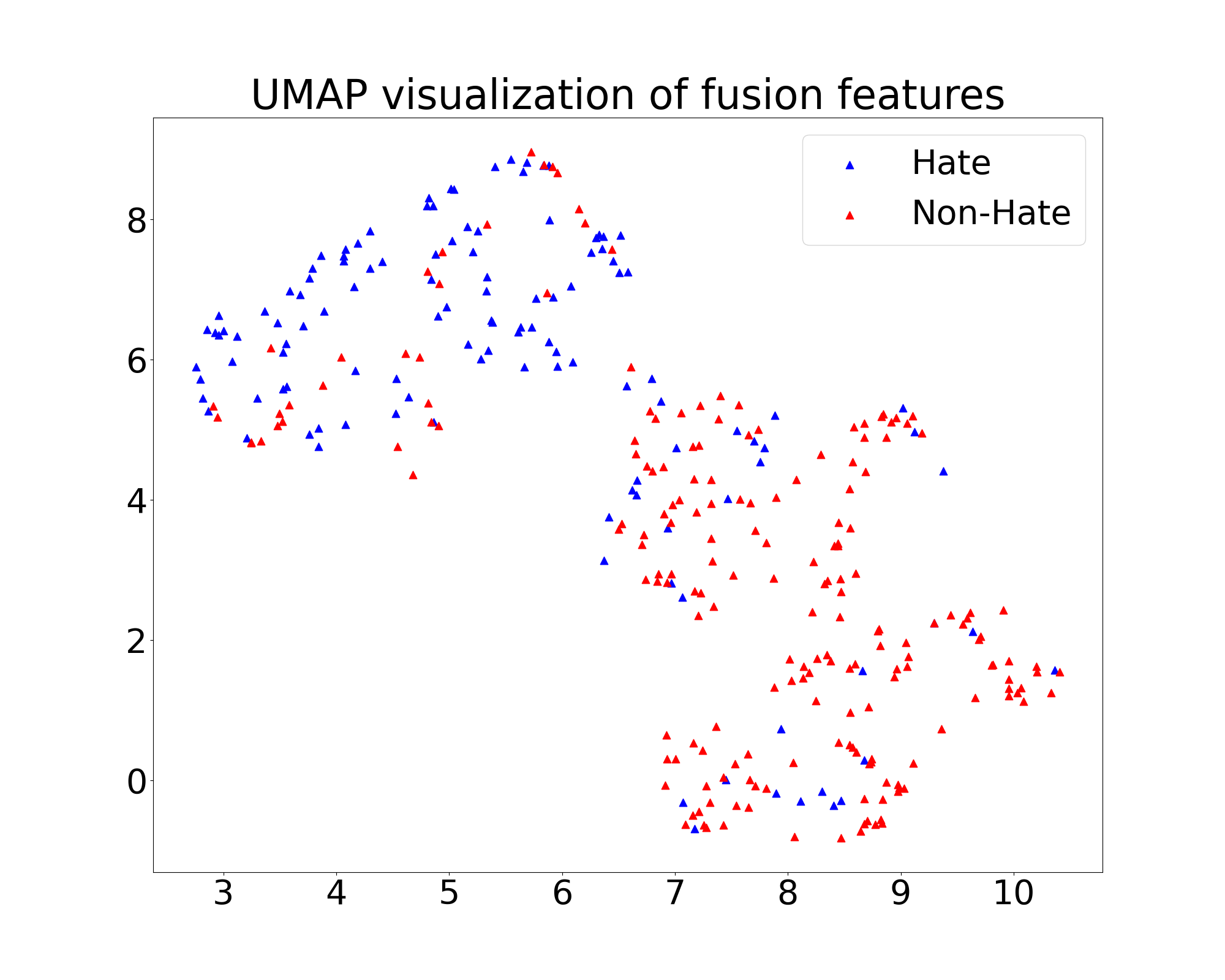}  
        \caption{Fusion with concatenation}
        \label{fig:hatemm1}
    \end{subfigure}
    \hfill
    \begin{subfigure}{0.49\linewidth}
        \centering
\includegraphics[width=\linewidth]{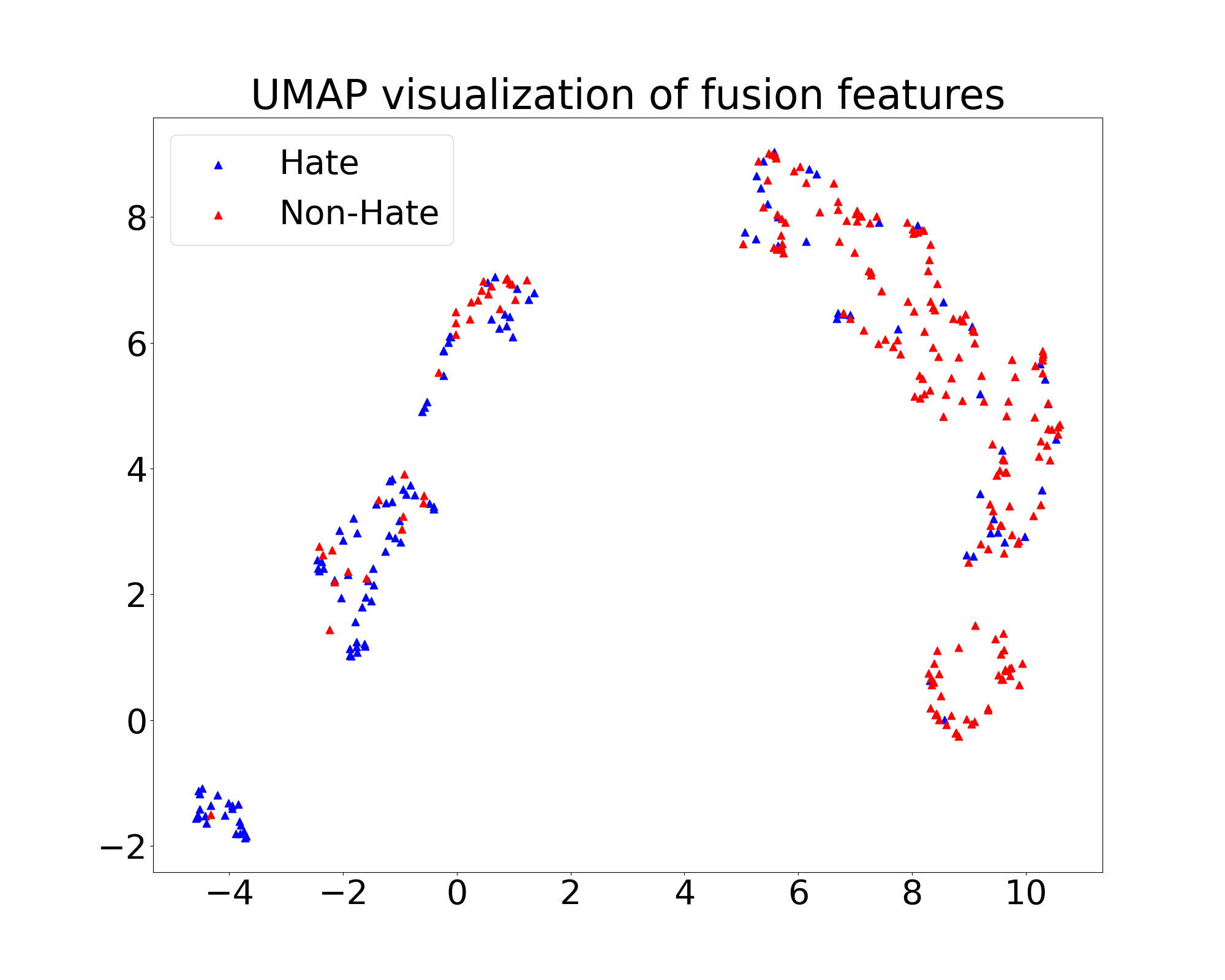}  
        \caption{CMFusion}
        \label{fig:CMFusion1}
    \end{subfigure}
    
    \caption{Comparison of feature visualisation with different fusion approaches. Subfigure (a) illustrates the feature visualisation of video samples in the HateMM dataset \cite{das2023hatemm}, where the features from video, text, and audio are fused using concatenation. In contrast, Subfigure (b) presents the feature visualisation of the same samples where the features from video, text, and audio are integrated through the CMFusion.}
    \label{fig:compare}
    \vspace{-1 em}
\end{figure}

Currently, much of the research on hate speech detection focuses on unimodal analysis, relying solely on text or images. However, unimodal approaches face inherent constraints as they depend on information from a single modality, making them inadequate in capturing complementary features or deep inter-modal correlations \cite{macavaney2019hate}. This results in a diminished capacity to effectively detection of hateful content \cite{badjatiya2017deep,del2017hate}. While multimodal approaches offer a more holistic perspective by incorporating multiple modalities, existing methods still fail to deliver efficient hate content detection. Some multimodal methods fail to encompass all relevant modalities, thereby failing to capture temporal dynamics and auditory components, both of which are critical for recognizing nuanced hateful content \cite{gonzalez2023understanding, kiela2020hateful}. A recent multimodal approach \cite{das2023hatemm} integrates text, audio, and video features simultaneously via a straightforward late fusion strategy, where features from different modalities are directly concatenated. However, as observed in Fig. \ref{fig:compare}(a), these approaches fail to exploit the complementary relationships between modalities, resulting in intertwined and indistinguishable feature spaces for hate and non-hate videos. This suggests a critical gap that calls for more comprehensive and effective solutions in hate video detection.

\begin{figure*}[t]
  \centering
  \includegraphics[width=0.95\textwidth]{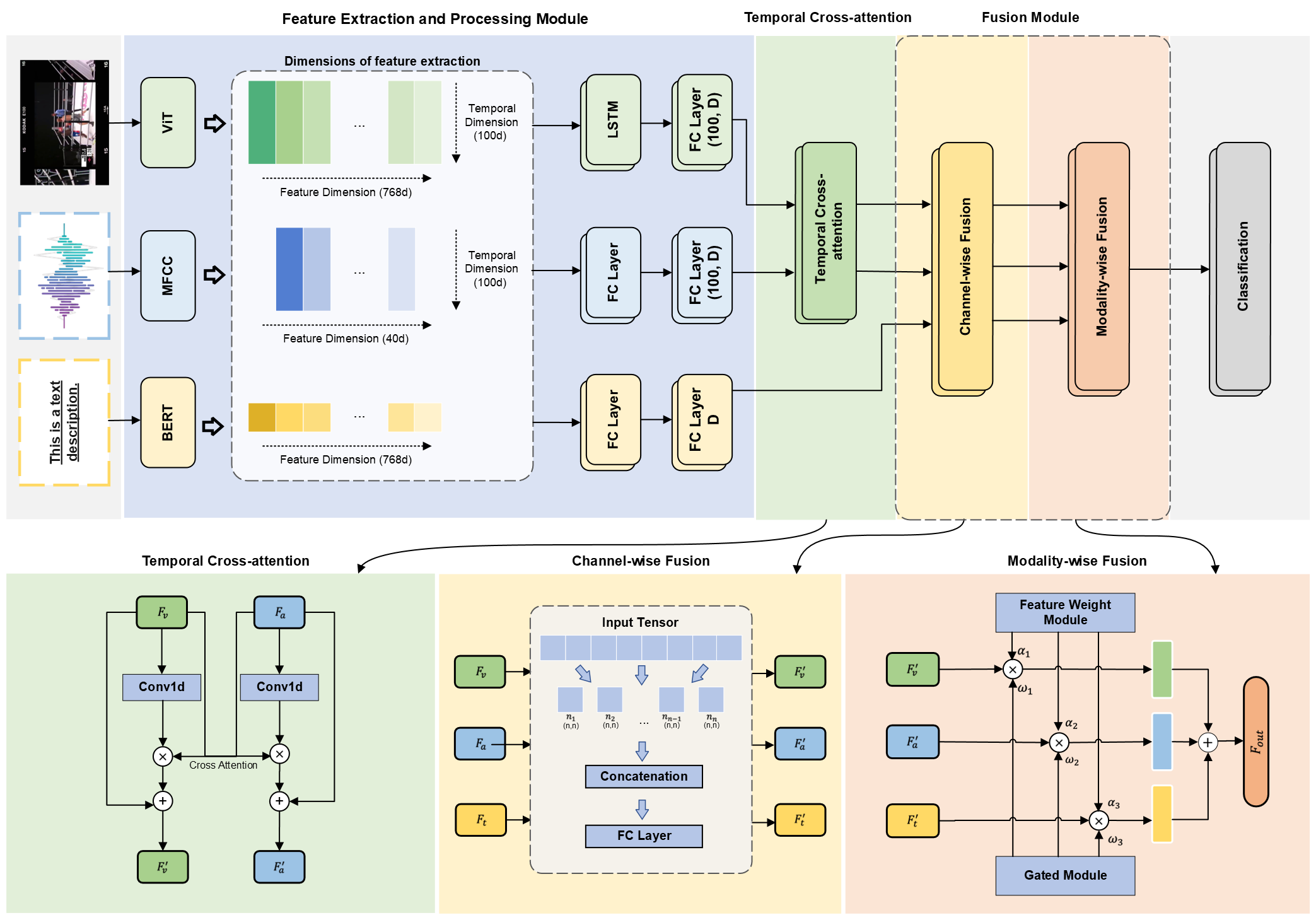} 
   \caption{The overview of the CMFusion Model along with the detailed structures of Channel-wise Fusion and Modality-wise Fusion. $F_v$, $F_a$, and $F_t$ represent the input vectors to each module, while $F'_v$, $F'_a$, and $F'_t$ represent the output vectors of the module.}

  \label{fig:4}
\end{figure*}

In this paper, we propose an enhanced multimodal hate video detection model based on \textbf{C}hannel-wise and \textbf{M}odality-wise \textbf{Fusion} Mechanism (CMFusion), as shown in Fig \ref{fig:4}.  CMFusion first utilizes pre-trained models to extract features from text, audio, and video data. Following this, a temporal cross-attention mechanism is applied between the video and audio modalities to capture their temporal dependencies. Furthermore, we integrate channel-wise and modality-wise fusion to seamlessly fuse the video, audio, and text features. The channel-wise fusion module aligns feature dimensions and enhances each modality’s representation, preparing for more effective fusion. The modality-wise fusion adaptively adjusts feature contributions, highlighting relevant information and improving synergy across modalities for better classification. As shown in Fig. \ref{fig:compare}(b), our fusion approach achieves clearer feature separation, demonstrating its effectiveness over Fig. \ref{fig:compare}(a). Our major contributions are:
\begin{itemize}
\item We present a new model that leverages temporal cross-attention mechanisms to facilitate interaction between video and audio modalities, enabling the capture of intricate features across different data types. This innovative approach effectively exploits the temporal correlations inherent in the video and audio streams, allowing the model to learn more nuanced and discriminative feature representations.
\item We exploit both channel-wise and modality-wise fusion techniques to better aggregate feature embeddings from individual modalities. The feature space visualization shows that our fusion module generates distinct representations, improving the separation between hate and non-hate content, which is crucial for effective hate video detection.
\item We demonstrate the superior performance of CMFusion on a real-world dataset\cite{das2023hatemm} over five widely-used baselines concerning accuracy, precision, recall, and F1 score. Further ablation studies, case studies and parameter analysis validate the overall design choices of CMFusion, reinforcing its capability in effectively detecting hate videos.

\end{itemize}
\section{Related work}
Numerous studies have investigated the classification of hate speech in textual data, with various approaches yielding promising results. MacAvaney et al. \cite{macavaney2019hate} proposed a multi-view Support Vector Machine (SVM) method for detecting hate speech, leveraging distinct feature views to enhance accuracy while offering greater interpretability of classification decisions. Similarly, Badjatiya et al. \cite{badjatiya2017deep} explored deep learning techniques such as FastText, Convolutional Neural Networks (CNNs), and Long Short-Term Memory networks (LSTMs)  for classifying tweets. Additionally, some studies have employed machine learning methods, such as SVM, for detecting hate speech\cite{del2017hate,sanoussi2022detection}.  However, these methods only consider a textual modality. Nuance and ambiguity in hate speech also present a significant challenge, as hate speech can be subtle or sarcastic, making it difficult for models to detect without a deep understanding of context or cultural nuances \cite{yin2021towards,rottger2020hatecheck,grondahl2018all}.

Yang et al. \cite{yang2019exploring} explored the classification of hate speech through multimodal fusion of text and images. Their study employed various fusion techniques, including simple feature concatenation, gated sum, bilinear transformation, and attention mechanisms. Kiela et al. \cite{lu2019vilbert} utilized the ViLBERT multimodal model, leveraging both visual and linguistic modalities to classify hate speech through the fusion of these two modalities \cite{kiela2020hateful}. The results indicated that combining image and text in multimodal approaches significantly improved the accuracy of hate speech detection \cite{kiela2020hateful}. Additionally, some studies employed spatial concatenation and attention mechanisms to fuse video and textual modalities \cite{gomez2020exploring,chhabra2023multimodal}, but they demonstrated that due to the complexity of multimodal interactions, models struggled to capture the intricate relationships between images and text.

Recently, Das et al. \cite{das2023hatemm} introduced the HateMM dataset, which annotates videos as either hateful or non-hateful. Their study explored the detection of hateful videos by integrating three modalities: video, text, and audio and compared the effectiveness of single-modality versus multi-modality approaches in identifying such content. In other classification tasks, such as emotion classification and sentiment analysis, multimodal features are typically fused using methods such as concatenation or element-wise summation \cite{xu2020social, delbrouck2020modulated,wang2022detecting}. Existing research indicates that even straightforward fusion methods enhance detection capabilities, as multimodal approaches provide a more comprehensive representation of the content compared to unimodal detection \cite{yang2012multi,bakkali2020visual,charte2018practical}. However, the fusion methods still require significant improvement to fully leverage the complementary strengths of different modalities for more effective hate video detection.
\section{Methodology}
This section outlines the framework of CMFusion, as illustrated in Fig. \ref{fig:4}. The CMFusion model comprises three primary modules: the Feature Extraction and Processing Module, the Temporal Cross-Attention Module, and the Channel-wise and Modality-wise Fusion Module. The Feature Extraction and Processing Module processes inputs from the three modalities—video, audio, and text—while maintaining their outputs as distinct modal features. The video and audio modalities are then fed into the Temporal Cross-Attention Module, which facilitates the capture of temporal dependencies between these streams. The outputs from this module, along with the text modality, are subsequently directed to the Channel-wise and Modality-wise Fusion Module, which optimizes the representation of information across all modalities. Finally, the integrated features generated by this module are utilized by the detection head for the multimodal detection task. This architecture ensures a comprehensive and synergistic approach to hate video detection, effectively leveraging the strengths of each modality.

\subsection{Problem Formulation}
The project defines the problem as follows: The multimodal task involves three modalities, video, audio, and text. Let \( F_n \) denote the feature vector of modality \( n \), where \( n \in \{v, a, t\} \) corresponds to the video (\( v \)), audio (\( a \)), and text (\( t \)) modalities, respectively. The feature vectors \( F_v \), \( F_a \), and \( F_t \) are processed through the Feature Extraction and Processing Module, the Temporal Cross-Attention Module, and the Channel-wise Fusion Module to obtain the modified features \( F_v' \), \( F_a' \), and \( F_t' \). These modified features are input into the Modality-wise Fusion Module to generate the weighted features \( FW_v \), \( FW_a \), and \( FW_t \). The final output feature is then combined as \( F_{\text{out}} = FW_v \oplus FW_a \oplus FW_t \), where $\oplus$ represents element-wise addition.
Our goal is to build a multimodal classifier \( C \) that maps the combined feature \( F_{\text{out}} \) to a binary label \( y \), that is, \( C: F_{\text{out}} \rightarrow y \), where \( y \in \{0, 1\} \) represents the label of the video content, with \( y = 0 \) indicating hate content and \( y = 1\) indicating non-hate content.

\subsection{Data pre-processing}For the audio and video modalities, we employ FFmpeg\cite{ffmpeg}, an open-source multimedia processing tool, to systematically extract audio data and frame images from each video. Specifically, we extract frames at consistent one-second intervals, yielding a total of 100 frames per video. 
If a video contains fewer than 100 frames, we use a white background to fill the remaining frame slots. Conversely, if a video exceeds 100 frames, we calculate the necessary step size based on the total fame count to ensure that exactly 100 evenly spaced frames are selected. This method allows us to analyze the video frame by frame while preserving the temporal characteristics of the video content.
After that, we employ the Whisper\cite{radford2023robust} developed by OpenAI to convert audio into analyzable text.
 
\subsection{Feature Extraction and Processing Module}
To extract visual features from video content, this study utilizes the Vision Transformer (ViT)\cite{dosovitskiy2020image} pre-trained model developed by Google. After preprocessing, each video yields 100 frames of image data. Each frame is individually processed through the ViT model to capture feature representations of the frame. Ultimately, the features from these 100 frames are aggregated into a large tensor with dimensions [100, 768], serving as the feature vector for the entire video. For the audio modality, this project employs the Mel Frequency Cepstral Coefficients (MFCC)\cite{muda2010voice} method to extract features, resulting in feature vectors with 40 dimensions. To synchronize with the temporal characteristics of the video, we extracted 100 time steps from the audio (padding with silence if shorter than 100 seconds, or downsampling with a fixed stride if longer). This process yields a final audio feature representation of size [100, 40]. For the text modality, a pre-trained BERT model is used to extract features from textual data, yielding feature vectors of 768 dimensions. 

In multimodal video detection tasks, capturing temporal
information is crucial, as the relevance of features can vary
significantly across different time points. To address this features from the video modality and audio modality are first input into an LSTM network, allowing for sequential preliminary processing of the original features extracted
from the video and audio. Subsequently, their dimensions are
standardized using a fully connected layer (FC Layer), which ensures
uniformity across all modalities, facilitating more effective
multimodal fusion and detection.  The specific operations
are as follows:
\begin{equation}
    f_{\text{FeatureProcess}}(F_x) = W \cdot \text{LSTM}(F_x^T) + b
\end{equation}
where \( x \) denotes the subset of input features for modality \( x \in \{v, a\} \), $f_{\text{FeatureProcess}}(F_x) \in \mathbb{R}^{\text{batch\_size} \times T \times D}$ is the final output of the model.

The text modality is processed using a simple neural network. The architecture consists of three fully connected layers, each followed by a ReLU activation function, as detailed below:
\begin{equation}
    f_{\text{FeatureProcess}}(F_t) = W_3 \cdot \text{ReLU}(W_2 \cdot \text{ReLU}(W_1 F_t + b_1) + b_2) + b_3
\end{equation}
where \(W_1\), \(W_2\), and \(W_3\) are the weight matrices of the three fully connected layers, and \(b_1\), \(b_2\), and \(b_3\) are the bias vectors of the three fully connected layers, respectively.

\subsection{Temporal Cross-attention Module}
Given that the video and audio modalities both contain temporal characteristics and can be aligned, we applied temporal cross-attention between these two modalities. This approach was inspired by the need to capture cross-modal dependencies over time, allowing the model to better understand synchronized patterns and interactions between the audio and visual sequences, ultimately improving its ability to recognize complex, time-dependent features. Specifically, we first apply 1D convolution (Conv1D) separately on the video features \( F_v \) and audio features \( F_a \) along the temporal dimension to extract sequential information. Following this, we perform temporal cross-processing between the two modalities, similar to the approach in SE-Net \cite{hu2018squeeze}, where the temporal attention weights are computed separately for each modality and used to facilitate cross-modal interactions. Given the input features \( F_v \) and \( F_a \), the module computes attention weights as follows:

\begin{equation}
C_x(t) =
\sum_{i=1}^{c_{\text{in}}} \sum_{k=1}^{K} F_x(t+k-1) \cdot w_i(k) + b,
\end{equation}
where \( {x} \) denotes the subset of input features for modality \( {x} \in \{v, a\} \). \( C_x(t) \) are the convolution output at time step \( t \); \( c_{\text{in}} \) is the number of input channels; \( F_m(t+k-1) \) is the input at time step \( t+k-1 \) and channel \( i \); \( w_i(k) \) is the weight of the convolutional kernel for channel \( i \) at the \( k \)-th position, with a kernel size of \( K \); \( b \) is the bias term; \( t \) is the current time step; and \( k \) is the index within the convolutional kernel's window.

After applying the temporal convolution for both video and audio modalities, we proceed with a cross-attention mechanism. Specifically, we apply the temporal cross-attention by using the audio temporal convolution on the video features and vice versa. The specific operations are as follows: 


\begin{equation}
F^{'}_v = F_v \times C_a + F_v, \quad F^{'}_a= F_a \times C_v + F_a
\end{equation}
This operation allows each modality to benefit from the temporal structure of the other, enhancing feature representations through temporal cross-attention.

\subsection{Fusion Module}
The fusion module consists of two parts, channel-wise fusion and modality-wise fusion. The channel-wise fusion module, structured as a multi-head linear layer, processes each modality's features independently, which can unify the dimensions of different modality features, ensuring that the fusion process is harmonized. The channel-wise fusion module is equipped with n heads, each with an \(n \times n\) linear layer to handle the corresponding segment of the input. The resulting feature segments are then concatenated into a single $D$-dimensional feature vector. Finally, an additional fully-connected layer is applied for feature enhancement. 
This configuration ensures that when this data is combined with data from other modalities, each modality contributes its most powerful information, thereby enhancing the efficiency and effectiveness of the final multimodal fusion. The three modalities pass through the channel-wise module, and their outputs are as follows:



\begin{equation}     
    \text{Output}_x = W_o \cdot [H_{x_0}, H_{x_1}, \ldots, H_{x_{\text{num\_heads}-1}}] + b_o
\end{equation}
where \( {x} \) denotes the subset of input features for modality \( {x} \in \{v, a, t\} \). \( W_o \) and \( b_o \) are the weight and bias, respectively. \( H_{x_i} \) denotes the transformation applied by the \(i\)-th head.

The Modality-wise Fusion module consists of a Feature Weight Module and a gated mechanism to adaptively select and highlight important parts of the input features for the current task or sample, ultimately fusing the features from the three modalities. 
Specifically, the feature scores are \(\alpha_i\ = \tanh(F_i W_a)v \), where \( F_i \in \{F_v, F_a, F_t\} \). \( W_a \) is a learnable feature weight matrix. \( v \) is the context vector. The gated scores \( w_i = \sigma(F_i W_g)\), where \( W_g \) are the weights of the gating linear layer. \( \sigma \) denotes the sigmoid function. 
Then, each modality feature can be calculated as follows,
\begin{equation}
F^{'}_{i} = \alpha_i \odot w_i \odot F_i, i \in \{v, a, t\},
\end{equation}
Finally, the fused feature \( F_{\text{fused}} \) is the direct summation of the calculated modality feature, \(F_{\text{fused}} = F^{'}_{v} + F^{'}_{a} + F^{'}_{t}\)
.



In the final stage of the model, the fused feature vector is classified through a linear layer combined with a SoftMax function to predict a classification score \(s = Softmax(\textbf{W}F_\text{fused})\), where \( s \in \mathcal{R}^{a}\) indicate the estimated probability of \(a\) classes in the dataset, and \textbf{W} denotes fully connected layers. 


\section{Experiments}

\subsection{Dataset and Experimental Settings}
The HateMM dataset \cite{das2023hatemm} used here contains 431 hate videos and 652 non-hate videos. Since the dataset is small, we used k-fold cross-validation for training and validation. The dataset is divided into 70\% for the training set, and 30\% for the test set, employing k-fold cross-validation. All experiments utilize a fixed data split with k set to 5. All experiments are conducted using PyTorch. The cross-entropy loss is used as the loss function. All models are trained using an initial learning rate of 1e-4 combined with the Adam optimizer. The batch size is set to 64 with a total of 40 epochs. Meanwhile, in the multi-head linear layer, the number of heads 
$n$ is set to 8. The dimensionality of the linear layer, $D$, is set to 64. 
All experiments are conducted on one NVIDIA L40 GPU. 
To assess the performance of CMFusion, we use the following evaluation metrics: Accuracy, Precision, Recall and F1 score. 
\subsection{Baseline}
To evaluate the effectiveness of the CMFusion model,
we compared it with five widely-used baselines, including both uni-modality methods and multi-modality methods. Specifically:
\begin{itemize}

\item\textbf{BERT}\cite{devlin2018bert}: Bidirectional Encoder Representations from Transformers, is a widely recognized transformer-based language model. In our experiments, we employed BERT to extract textual features from transcripts of audio and subsequently utilized these features for detection.
\item\textbf{GPT-3.5\footnote{For simplicity, we use GPT-3.5 to refer to GPT-3.5 Turbo.}}\cite{brown2020language}: GPT-3.5  is a transformer-based autoregressive language model renowned for its ability to generate and comprehend text at scale. We employ GPT-3.5 as a baseline model due to its state-of-the-art performance in natural language understanding and generation tasks. Specifically, during our interaction with GPT-3.5, we provided the following preset prompt: "\textit{Please determine whether the following text contains hateful content. If it contains hateful content, please return 0; if it does not contain hateful content, please return 1.}" The model's output, in the form of a predicted label, is then compared with the true label for evaluation. While GPT-3.5 is not specifically optimized for hate speech detection, its ability to capture complex linguistic patterns makes it a strong candidate as a baseline.
\item\textbf{ViT}\cite{dosovitskiy2020image}: ViT (Vision Transformer) is a model proposed by Google's research team for computer vision tasks. We employ the Vision Transformer (ViT) to extract features from video frames. These features are subsequently passed through a Long Short-Term Memory (LSTM) network and fully connected layers, followed by a classifier to obtain the detection result for the video modality.

\item \textbf{MFCC}\cite{muda2010voice}: Mel-frequency Cepstral Coefficients is a widely used feature extraction technique in speech processing and recognition. We use MFCC to extract audio features. The extracted audio features are processed through an LSTM network and fully connected layers, followed by a classifier to obtain the detection result for the audio modality.
\item\textbf{HateMM}\cite{das2023hatemm}: We selected HateMM as one of our baselines because it is currently the only dataset focused on multimodal hate speech detection in videos, making it a crucial resource for this domain. Our project also builds upon this dataset to develop our model. HateMM's model architecture combines features from three modalities—text, video, and audio. These features are then fused using a multimodal fusion layer, which integrates information from all modalities before being passed through dense layers for detection.

\end{itemize}

\subsection{Overall Results}
\begin{table}[t]
\centering
\caption{Performance Comparison of CMFusion and Five Baseline Models (Best results in bold).}
\label{tab:my-table1}
\scriptsize
\begin{tabular}{c|c|c|c|c|c}
\hline
\textbf{Modality} & \textbf{Model} & \textbf{Accuracy} & \textbf{F1 score} & \textbf{Precision} & \textbf{Recall} \\ \hline
T                 & BERT           & 0.798             & 0.838             & 0.81               & 0.868           \\
T                 & GPT-3.5        & 0.663             & 0.647             & \textbf{0.854}     & 0.521           \\
V                 & ViT            & 0.756             & 0.806             & 0.774              & 0.842           \\
A                 & MFCC           & 0.710             & 0.767             & 0.745              & 0.793           \\
V+A+T             & HateMM         & 0.803             & 0.841             & 0.811              & 0.874           \\
\textbf{V+A+T}    & \textbf{CMFusion} & \textbf{0.823} & \textbf{0.860}     & 0.817              & \textbf{0.908}  \\ \hline
\end{tabular}
\end{table}
Table I shows the comparison between the CMFusion model and the five baseline models. We can see that the CMFusion model achieves the best results across nearly all evaluation metrics, particularly in F1-score and Recall. This indicates the model’s ability to effectively capture multimodal information, thereby enhancing detection accuracy.

Compared to single-modality models such as BERT and ViT, as well as the text-only GPT-3.5 model, CMFusion demonstrates a significant advantage in performance through the integration of text, video, and audio modalities. Although GPT-3.5 performs slightly better in Precision, its performance in other metrics, especially Recall, falls far behind CMFusion, indicating that GPT-3.5 might suffer from a trade-off between high precision and low recall. Meanwhile, BERT, ViT, and MFCC are constrained by their single-modality approach, resulting in overall performance that lags behind CMFusion.

When comparing HateMM with our proposed CMFusion model, CMFusion consistently outperforms HateMM across all key metrics. Specifically, CMFusion achieves an accuracy of 0.823, which represents a 2.49\% improvement over HateMM's 0.803. This enhancement is attributed to CMFusion's more advanced multimodal fusion mechanism, which effectively integrates and synchronizes information from video, audio, and text modalities. CMFusion's F1 score is 0.86, clearly surpassing HateMM's 0.841. The F1 score is particularly important for hate video detection tasks because it ensures that the model not only correctly identifies hateful content but also minimizes the chances of missing such content. The precision of both models is relatively close, with CMFusion scoring 0.817 and HateMM at 0.811. In terms of recall, CMFusion's performance is outstanding, reaching 0.908, a 3.89\% increase over HateMM's 0.874. The improvement in recall is particularly significant in real-world applications. For example, on a social media platform, if a piece of hateful content goes undetected (i.e., a false negative occurs), the platform may face public backlash or even legal consequences. Therefore, a higher recall helps the platform more comprehensively detect and remove hateful content, contributing to a safer community environment.


\begin{figure*}[t]  
    \centering
    \begin{subfigure}{0.32\linewidth}  
        \centering
        \includegraphics[width=\linewidth]{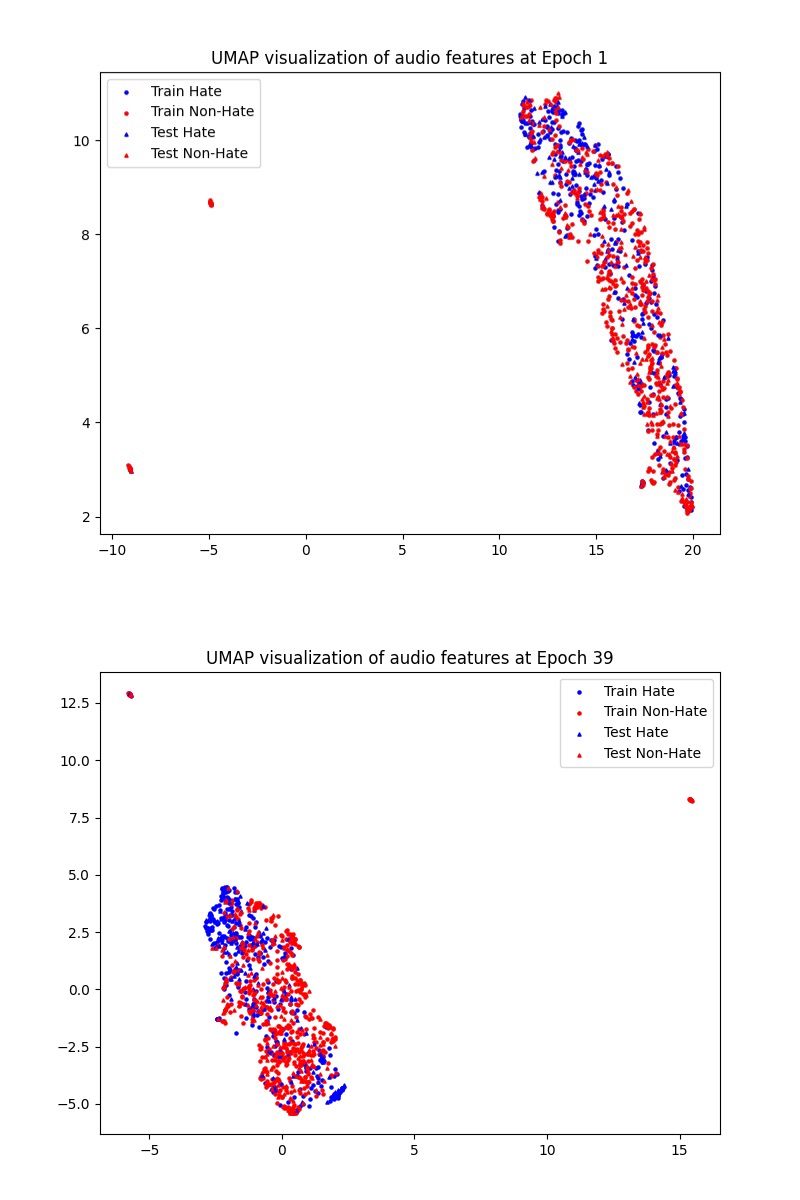}  
        \caption{Audio Feature}
        \label{fig:audio3}
    \end{subfigure}
    \begin{subfigure}{0.32\linewidth}
        \centering
        \includegraphics[width=\linewidth]{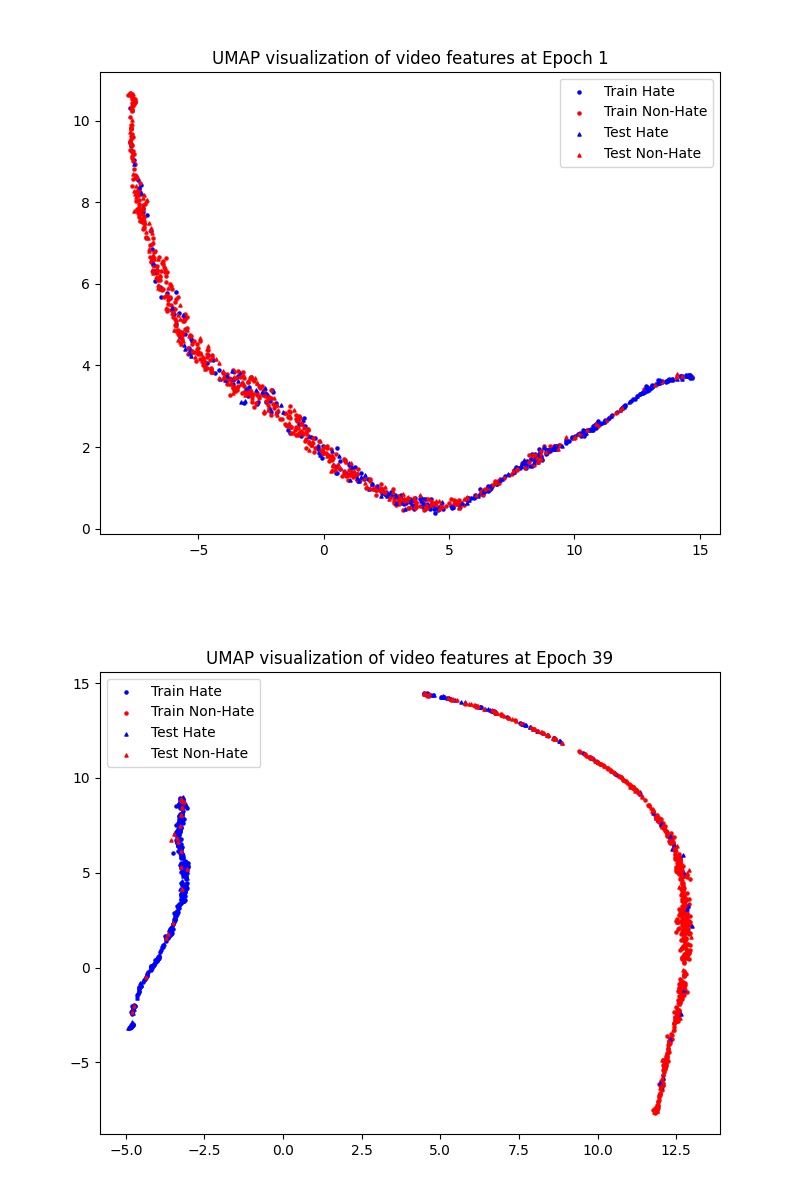}  
        \caption{Video Feature}
        \label{fig:video3}
    \end{subfigure}
    \begin{subfigure}{0.32\linewidth}  
        \centering
        \includegraphics[width=\linewidth]{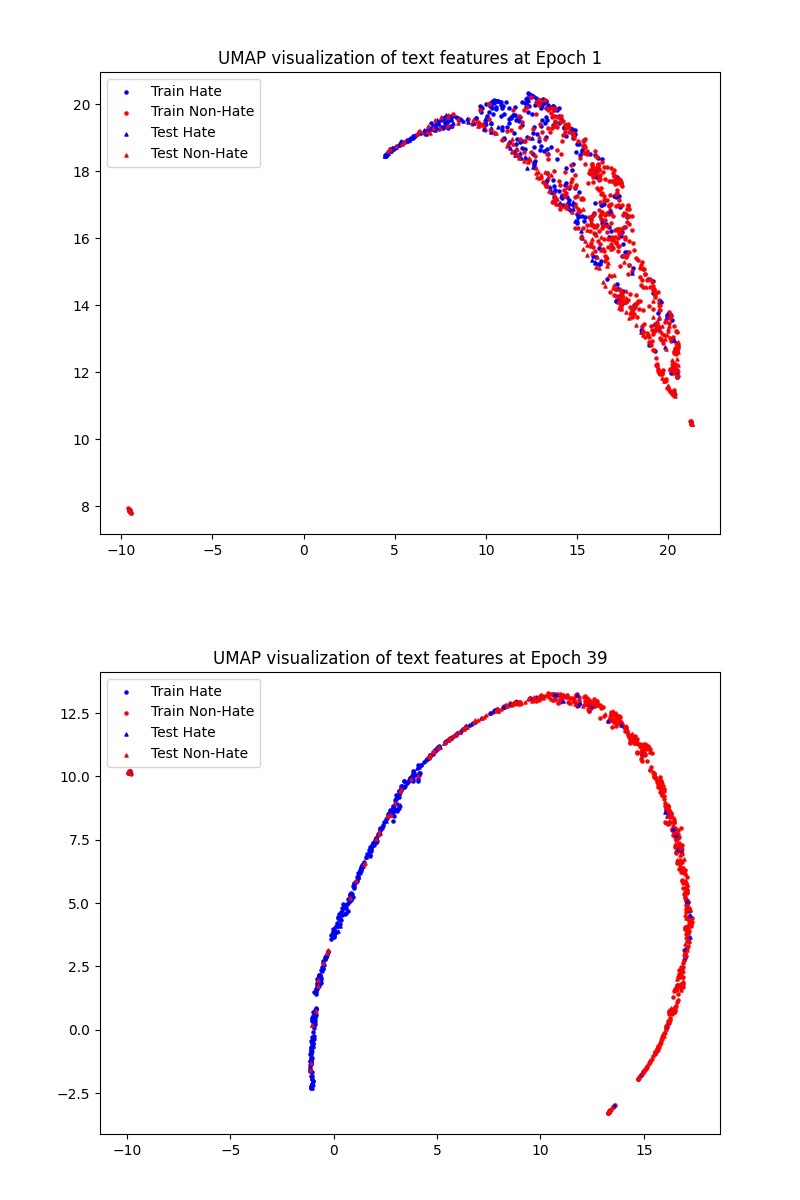}  
        \caption{Text feature}
        \label{fig:text3}
    \end{subfigure}
    
    \caption{UMAP visualizations of the feature representations from audio, video, and text modalities, respectively.}
    \label{fig:umap_video}
\end{figure*}
\subsection{Ablation Study}
Table II presents the results of the ablation experiments, where different fusion strategies and architectures are compared. These experiments allow us to analyze the impact of various components in the proposed CMFusion model, particularly the effects of concatenation (\(\odot\)), element-wise addition (\(\oplus\)), Temporal Cross Attention (TCA), and Channel-wise and Modality-wise Fusion (CMF).
\begin{table}[t]
\caption{Ablation Experimental Results. $\odot$: Represents the concatenation operation. $\oplus$: Represents element-wise addition. TCA: Denotes Temporal Cross Attention. CMF: Denotes Channel-wise and Modality-wise Fusion. }
\label{tab:my-table}
\resizebox{\columnwidth}{!}{%
\begin{tabular}{
>{\columncolor[HTML]{FFFFFF}}c |
>{\columncolor[HTML]{FFFFFF}}c 
>{\columncolor[HTML]{FFFFFF}}c 
>{\columncolor[HTML]{FFFFFF}}c 
>{\columncolor[HTML]{FFFFFF}}c }
\hline
\textbf{Architecture}     & \textbf{Accuracy}    & \textbf{F1 score}   & \textbf{Precision}   & \textbf{Recall}      \\ \hline
V$\odot$A               & 0.760  & 0.813 & 0.765 & 0.871 \\
V$\odot$T               & 0.803 & 0.843 & 0.809 & 0.882 \\
T$\odot$A               & 0.813 & 0.850  & 0.820  & 0.883 \\
V$\odot$A$\odot$T      & 0.809 & 0.847 & 0.816 & 0.883 \\
(V$\odot$A$\odot$T)$^{\text{TCA}}$ & 0.785 & 0.827 & 0.798 & 0.861 \\
(V$\oplus$A$\oplus$T)$^{\text{CMF}}$               & 0.810  & 0.850  & 0.807 & 0.900   \\
\textbf{(V$\oplus$A$\oplus$T)}$^{\text{CMFusion}}$
 & \textbf{0.823} & \textbf{0.860} & \textbf{0.817} & \textbf{0.908} \\ \hline
\end{tabular}%
}
\end{table}

\textbf{Two-Modality Fusion:} The fusion of two modalities, such as video (V) and audio (A) or text (T) and audio (A), demonstrates reasonable performance. The V\(\odot\)A fusion achieves an accuracy of 0.76 with an F1 score of 0.813, while T\(\odot\)A achieves an F1 score of 0.85 and a precision of 0.82. This indicates that bimodal fusion can capture useful information, but lacks the comprehensive integration provided by three-modality fusion.

\textbf{Three-Modality Fusion:} The combination of V\(\odot\)A\(\odot\)T through concatenation results in significant improvement across all metrics, reaching an accuracy of 0.809 and an F1 score of 0.847. This demonstrates that integrating information from all three modalities—video, audio, and text—provides a considerable advantage over using only two modalities.
    
\textbf{Impact of Temporal Cross Attention (TCA):} 
Introducing Temporal Cross Attention (TCA) helps improve the temporal alignment between modalities. The (V$\odot$A$\odot$T)$^{\text{TCA}}$ model achieves an F1 score of 0.827 but shows a slight decrease in precision. More importantly, the accuracy remains at 0.785, indicating that the temporal alignment achieved by TCA alone does not lead to a significant improvement in overall accuracy. This suggests that while temporal alignment contributes to better synchronization between modalities, further enhancements, such as more advanced fusion methods, are needed to fully exploit the potential of multimodal integration.

\textbf{Channel-wise and Modality-wise Fusion (CMF):} The (V$\oplus$A$\oplus$T)$^{\text{CMF}}$ model, which incorporates Channel-wise and Modality-wise Fusion, performs well, achieving an F1 score of 0.85 and a recall of 0.9. This highlights its ability to effectively integrate multimodal information and capture a broader range of positive instances.

\textbf{CMFusion:} The proposed CMFusion model, which utilizes both temporal and modality fusion mechanisms, outperforms all other configurations. With an accuracy of 0.823, an F1 score of 0.86, and a recall of 0.908. This highlights the importance of combining element-wise addition for better feature integration, along with sophisticated temporal and modality fusion mechanisms, to maximize performance.

\subsection{Case Study I Fusion Approach Analysis}
In this section, we analyze four fusion methods and compare their results, as presented in Table III.

\begin{table}[t]
\centering
\caption{Performance Comparison of Different Fusion Methods.}
\label{tab:my-table2}
\footnotesize 
\begin{tabular}{c|cccc}
\hline
\textbf{Model} & \textbf{Accuracy} & \textbf{F1 score} & \textbf{Precision} & \textbf{Recall} \\ \hline
M1             & 0.782             & 0.830             & 0.782              & 0.884           \\
M2             & 0.803             & 0.839             & 0.820              & 0.861           \\
M3             & 0.811             & 0.850             & 0.810              & 0.894           \\
\textbf{M4}    & \textbf{0.816}    & \textbf{0.854}    & \textbf{0.815}     & \textbf{0.897}  \\ \hline
\end{tabular}
\end{table}

%
%
%

\textbf{Temporal Cross Attention with Sum Fusion (M1): }
This study explores the use of temporal cross attention between the video and audio modalities, followed by element-wise sum fusion of the three modalities. This method specifically emphasizes the interaction between temporal information and modality relationships. By applying temporal cross attention to the video and audio modalities and using sum fusion, the goal is to capture cross-modal interactions over different time periods. By focusing on temporal interactions, the aim is for M1 to better capture time-synchronized information between modalities compared to simpler sum or concatenation methods. This approach is particularly relevant in multimodal sequential tasks, such as detecting long-duration hateful speech or behaviours.

\textbf{Modality Concatenation Prior to Dense Layer (M2): }
In this method, the features from the three modalities are concatenated first and then passed through a dense layer. This approach does not impose any assumptions about the relationships between modalities, allowing the dense layer to learn the fusion weights and interactions autonomously. It is a relatively straightforward and simple fusion method. M2 is used to evaluate the effectiveness of modality fusion without complex interaction mechanisms, providing a baseline for comparison against more sophisticated fusion strategies.

\textbf{Modality-Specific Weights Prior to Concatenation (M3): }
This method assigns specific, trainable weights to each modality, aiming to give greater influence to certain modalities. These weights are trainable, allowing the model to automatically adjust the importance of each modality based on the task. M3 focuses on the contribution of each modality to the detection task. Since the contribution of different modalities can vary across tasks, the inclusion of modality-specific trainable weights enables the model to more flexibly handle imbalanced modality contributions.

\textbf{Temporal Attention with Channel-wise and Modality-wise Fusion before Concatenation (M4):}
The M4 model differs from CMFusion in that CMFusion uses sum fusion, while M4 uses concatenation for fusion. This study aims to explore the effects of concatenation versus sum fusion. In some multimodal fusion tasks, concatenation often yields better results, but in our task, sum fusion has demonstrated superior performance. This may be due to the fact that sum fusion results in lower feature dimensionality, which simplifies the model and reduces its complexity.

\subsection{Case Study II Feature Space Analysis}
Fig. 3 presents UMAP\cite{mcinnes2018umap} visualizations of the feature representations from audio, video, and text modalities, respectively, at different training epochs. These visualizations provide insight into how well the model learns to separate hate and non-hate content within each feature space as training and testing progress. Across all three modalities, we observe an increasing separation between hate and non-hate classes over time, with the most distinct separation occurring in later epochs. However, the degree of separation varies between modalities. Specifically, while the separation of audio features between hate and non-hate samples improves over time, the distinction remains less pronounced compared to the video and text modalities. This suggests that the audio modality may contribute useful but somewhat less discriminative features for this specific detection task. Video features and text features show a clearer and more structured separation of hate and non-hate content, particularly at epoch 39.  The UMAP visualization reveals well-defined clusters, suggesting that video and text features provide significant cues for the model to differentiate between the two classes.

\section{Conclusion}
In this work, we introduce a novel CMFusion framework for multimodal hateful content detection.  Our model can first learn the temporal correlations via a temporal cross-attention mechanism between video and audio modalities for robust temporal modelling.
Then, by integrating channel and modality-wise fusion, our CMFusion can effectively align different modalities of feature and learn corresponding adaptive weights for better classification.
Extensive experiments demonstrated the superiority of our model, surpassing five widely-used baselines, including uni-modality models, LLMs, and multi-modality models. Further ablation studies and case studies validate the effectiveness of our model design and highlight the potential of advanced feature fusion methods in enhancing hate video detection.

\section*{Declaration}

This paper may contain references to sensitive content related to hate speech. However, it is presented solely for research purposes and does not intend to promote, propagate, or endorse any form of hate, discrimination, or harmful behaviour.

\section*{Acknowledgment}
This work was supported by the Alan Turing Institute and DSO National Laboratories Framework Grant Funding. We are grateful to Jianbo Jiao, Qingyue Sun, and Fu Wang for their valuable insights and suggestions on this work.

\bibliography{main}

\end{document}